\documentstyle[12pt,blois,epsf]{article}

\begin{document}
\heading{EVOLUTION OF ELLIPTICAL GALAXIES\\ 
         IN A HIERARCHICAL UNIVERSE}

\author{G. Kauffmann $^{1}$, S. Charlot $^{2}$} { $^{1}$ Max Planck Institut f\"{u}r Astrophysik, Munich, Germany} {$^{2}$ Institut D'Astrophysique, Paris, France}

\begin{bloisabstract}
In hierarchical models of galaxy formation, ellipticals form from the merging
of disk galaxies drawn together by gravity as their surrounding dark halos coalesce.
Using semi-analytic techniques, we are able to follow  merging, star 
formation and chemical enrichment in galaxies as a function of redshift
for a given set of cosmological initial conditions.
We calculate the ages and metallicities of cluster elliptical galaxies
formed by mergers in a standard CDM cosmology and show that we can reproduce
the observed correlation between absolute magnitude and colour, and the scatter about this
relation.
We also study the evolution of the elliptical population to high redshift.
In rich clusters, the models predict  a substantial population of ``evolved'' ellipticals
at $z>1$. 
However, the {\em average number density} of bright ellipticals decreases at high redshift.
The rate of decrease is sensitive to cosmological parameters.
In  high-density CDM models, the abundance of bright ellipticals declines by a factor
of 3 by $z=1$. In a low-density CDM  model ($\Omega=0.3$, $\Lambda=0.7$), the abundance
has only dropped by 30\% by $z=1$, but is down by a factor 3 at $z=2$. 
\end{bloisabstract}

\section{Brief Description of the Models}
The semi-analytic models used in this paper have been described in detail in
\cite{KWG}, \cite{KC} and most recently, in \cite{KD}. The reader is referred to these
papers for more background. 
To summarize,
an algorithm based on an extension of the Press-Schechter theory is used to generate Monte
Carlo realizations of the merging paths of dark matter halos from high
redshift until the present. 
It is assumed that as a dark matter halo forms, the gas relaxes to a 
distribution that exactly 
parallels that of the dark matter.  
Gas then cools and condenses onto a central galaxy at the core of each
halo. Star formation, feedback processes and chemical enrichment 
take place as described in \cite{KWG} and 
\cite {KC}.       
As time proceeds, a halo will
merge with a number of others, forming a new halo of larger
mass. All gas which has not already cooled is assumed to be shock
heated to the virial temperature of this new halo. This hot gas then
cools onto the central galaxy of the new halo, which is identified
with the central galaxy of its {\em largest progenitor}. The central
galaxies of the other progenitors become {\em satellite galaxies}, 
which are able to merge with the central galaxy on a dynamical friction
timescale. If  a merger takes place between two galaxies of roughly comparable
mass, the merger remnant is labelled as an elliptical and all cold gas
is transformed instantaneously into stars in a ``starburst''. 
Note that the infall of new gas onto  
satellite galaxies is not allowed, and star
formation will continue in such objects only until their existing cold gas
reservoirs are exhausted. Thus the epoch at which a galaxy is accreted by a larger
halo delineates the transition between active star formation in the galaxy
and passive evolution of its stellar population. The stellar populations of 
elliptical merger remnants in clusters hence redden as their stellar populations age.
Central galaxy merger remnants in the ``field'' are able to accrete new gas in the form of a disk 
to form a spiral galaxy consisting of both a spheroidal bulge and a disk
component. We use the metallicity-dependent  population synthesis models
of Bruzual \& Charlot \cite{BC} to calculate observable quantities such as magnitudes and colours for our
model galaxies.
Note that early-type galaxies are defined to  to have $M(B)_{bulge}-M(B)_{total} < 1$ 
\cite{SV}.

\section {Ages, metallicities and the evolution of the colour-magnitude relation}
In figure 1, we compare the age-luminosity and metallicity-luminosity relations of cluster and field   
elliptical galaxies at the present day. 
Field ellipticals are defined to reside in halos with $V_c < 600$ km s$^{-1}$
(i.e. systems with line-of-sight velocity dispersion less than about 400 km s$^{-1}$).
Results are shown for a ``standard'' CDM cosmology ($\Omega=1$, $\Gamma=0.5$, $\sigma_8=0.7$
and $H_0=50$ km s$^{-1}$ Mpc$^{-1}$).
Elliptical galaxies have V-light weighted mean ages of $\sim 8$ Gyr, with an r.m.s.
scatter of $\sim 1$ Gyr. The metallicities range from 0.5 solar at M(V)= --18
to just over solar for the brightest systems in the cluster.

As can be seen, the relations between metallicity and luminosity are identical for field and
cluster Es. The bright field ellipticals have systematically younger mean stellar ages
and have an r.m.s. scatter in age about 60\% larger than ellipticals of the same luminosity
in clusters. This is because the bright field Es are almost always ``central'' galaxies 
that have formed in a recent merger and have not yet had time to accrete
a new disk component.

\begin{figure}
\centerline{
\epsfxsize=9cm \epsfbox{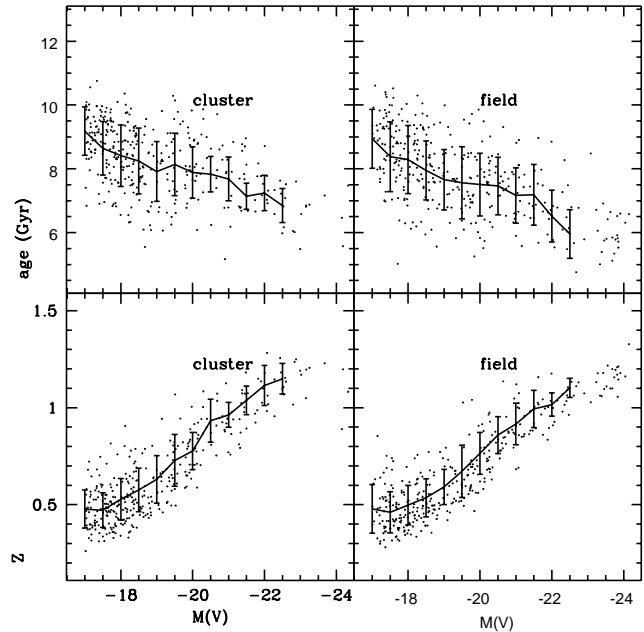}
}
\caption{\label{fig1}
The metallicity-luminosity and age-luminosity relations of cluster ellipticals 
are compared with the relatiions for field ellipticals.V-light weighted mean
quantities are plotted for each galaxy.
The solid line shows the mean relation and the error bars are one sigma standard deviations}
\end {figure}

In figure 2 we show the predicted  restframe $U-V$ colour-magnitude
relation for early-type galaxies in clusters with $V_c= 1000$ km s$^{-1}$
at $z=0, 0.7, 1.5$ and $2$.                                                         
For reference, the solid line is the observed relation of Bower, Lucey \& Ellis 
\cite{BLE} at $z=0$.
The scatter in the model relation at $z=0$ is 0.04 mag, which is in good agreement with
the observations. There is no noticeable increase in the scatter at higher redshifts.
Note, however,  that there is a {\em progressive flattening} 
in the slope of the colour-magnitude 
relation with redshift, although it only becomes really noticeable at $z > 1$.
This is caused by the fact that the bigger ellipticals have slightly younger mean ages
in our model, and the effect of this on  observables such as colour and magnitude, is much
more pronounced at early times.

\begin{figure}
\centerline{
\epsfxsize=9cm \epsfbox{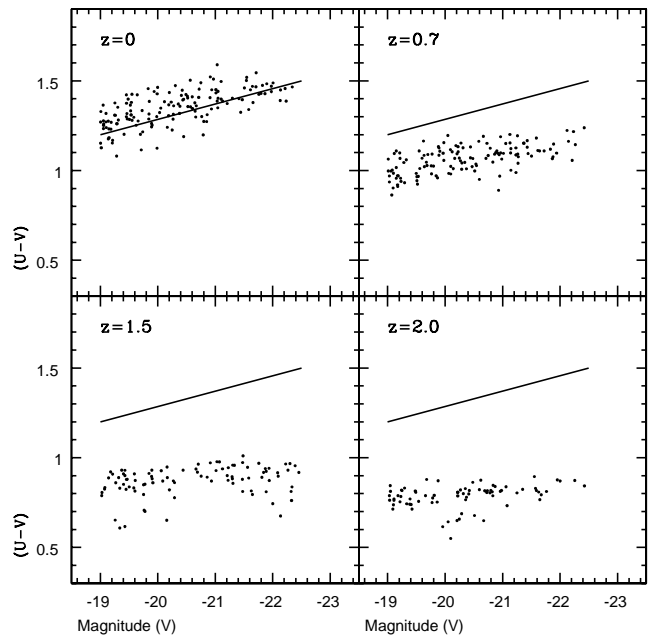}
}
\caption{\label{fig2}
The rest-frame $U-V$ colour-magnitude relation of cluster  elliptical galaxies 
at z=0.0.7, 1.5 and 2.}
\end {figure}

\section {Abundance Evolution} 
In figure 3, we show the predicted evolution with redshift of the 
abundance of elliptical galaxies
with stellar mass greater than $10^{11} M_{\odot}$ for a high-density $\tau$CDM model
($\Omega=1$, $\Gamma=0.2$, $\sigma_8=0.6$, $H_0 = 50$ km s$^{-1}$ Mpc$^{-1}$) and for 
a low-density $\Lambda$CDM
model ($\Omega=0.3$, $\Lambda=0.7$, $\sigma_8=0.9$, $H_0= 70$ km s$^{-1}$ Mpc $^{-1}$).
As expected, the abundance of elliptical galaxies decreases with redshift in both models,
because these systems are formed continuously over time by merging. The
rate of evolution is  more rapid in the high-density model. 
For $\tau$CDM, the abundance of bright ellipticals declines by a factor
of 3 by $z=1$. For $\Lambda$CDM, the abundance
has only dropped by 30\% by $z=1$, but is down by a factor 3 at $z=2$. 
This simply reflects the fact that structure formation occurs {\em later}
in the high-density model than in the low-density one.

As discussed in \cite{KC2}, future wide area $K$-selected redshift surveys will
enable the buildup of elliptical galaxies through mergers to be quantified more
accurately.

\begin{figure}
\centerline{
\epsfxsize=10cm \epsfbox{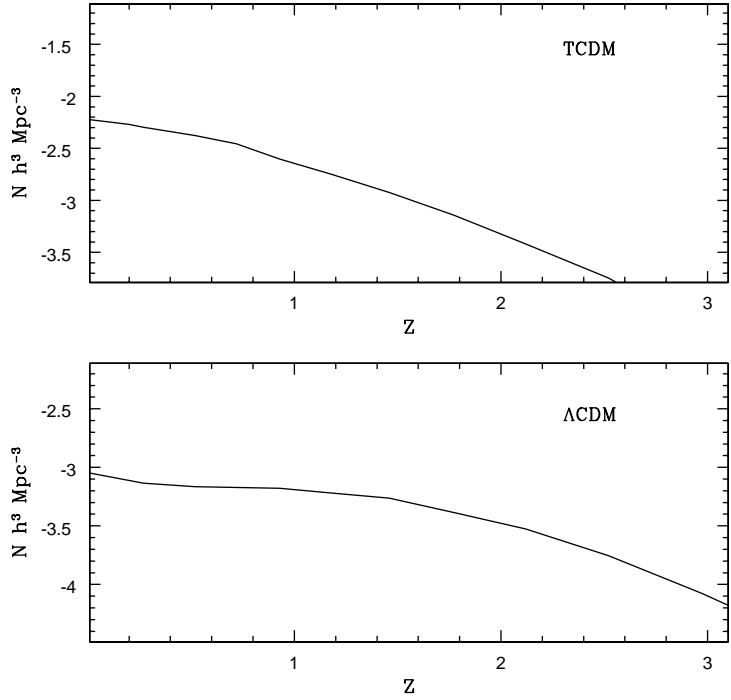}
}
\caption{\label{fig3}
The evolution of the co-moving abundance of elliptical galaxies with stellar
mass greater than $10^{11} M_{\odot}$ in a high-density ($\tau$CDM) and low-density
($\Lambda$CDM) model.}
\end {figure}

\vfill
\end{document}